\documentclass[]{aa}
\usepackage{graphicx}
\usepackage{color}
\usepackage{url}
\usepackage{amsmath}

\begin{document}

\title{Prominence instability and CMEs triggered by massive coronal rain in the solar atmosphere}

\author{Vashalomidze, Z.\inst{1,2,3}, Zaqarashvili, T.V.\inst{1,2,3}, Kukhianidze, V.\inst{2,3}, Ramishvili, G.\inst{2,3}, Hanslmeier, A.\inst{1}, G\"om\"ory, P.\inst{4}
}

 \institute{IGAM, Institute of Physics, University of Graz, Universit\"atsplatz 5, 8010 Graz, Austria\\
                                \and
            Ilia State University, Kakutsa Cholokashvili Ave 3/5, 0162 Tbilisi, Georgia  \\
            \email{zurab.vashalomidze.1@iliauni.edu.ge}
                                \and
        E.Kharadze Georgian National Astrophysical Observatory, Mount Kanobili, Abastumani, Georgia \\
                                \and
        Astronomical Institute, Slovak Academy of Sciences, P.O. Box 18, 05960 Tatransk\'a Lomnica, Slovak Republic\\
 }

\date{Received / Accepted }

\abstract{Triggering process for prominence instability and consequent CMEs is not fully understood. Prominences are maintained by the Lorentz force against the gravity, therefore reduction of the prominence mass due to the coronal rain may cause the change of the force balance and hence destabilisation of the structures.}
{We aim to study the observational evidence of the influence of coronal rain on the stability of prominence and subsequent eruption of CMEs.}
{We used the simultaneous observations from AIA/SDO and SECCHI/STEREO spacecrafts from different angles to follow the dynamics of prominence/filaments and to study the role of coronal rain in their destabilisation. }
{Three different prominences/filaments observed during years 2011-2012 were analysed using observations acquired by SDO and STEREO. In all three cases massive coronal rain from the prominence body led to the destabilisation of prominence and subsequently to the eruption of CMEs. The upward rising of prominences consisted in the slow and the fast rise phases. The coronal rain triggered the initial slow rise of prominences, which led to the final instability (the fast rise phase) after 18-28 hours in all cases. The estimated mass flux carried by coronal rain blobs showed that  the prominences became unstable after 40 \% of mass loss.}
{We suggest that the initial slow rise phase was triggered by the mass loss of prominence due to massive coronal rain, while the fast rise phase, i.e. the consequent instability of prominences, was caused by the torus instability and/or magnetic reconnection with overlying coronal field. Therefore, the coronal rain triggered the instability of prominences and consequent CMEs. If this is the case, then the coronal rain can be used to predict the CMEs and hence to improve the space weather predictions. }


\titlerunning{Prominence instability and CME Eruption}

\authorrunning{Vashalomidze et al.}

\maketitle

\section{Introduction}

Solar prominences/filaments are relatively cool and dense structures in the tenuous, hot solar corona (Labrosse et al. \cite{Labrose2010}; Arregui et al. \cite{Arregui2012}). Most of active region prominences are relatively unstable and lead to coronal mass ejections (CME), which affect space weather conditions near the Earth (Panesar et al. \cite{Panesar2014}; Schmieder et al. \cite{Schmieder2002}, Gopalswamy et al. \cite{Gopalswamy2003}). Prominence instability, the initiation of CMEs (Priest et al. \cite{Priest2002}), as well as an interconnection between CMEs and erupting prominences are not clearly understood (Chae et al. \cite{Chae2000}, Zhang et al. \cite{Zhang20017a}, Zirker et al. \cite{Zirker1998}). Observations show that the prominences/filaments are supported by the coronal magnetic field against the gravity (Ning et al. \cite{Ning2009b}, Shen et al. \cite{Shen2015}, Zhang et al. \cite{Zhang20017b}). Therefore, some process (or processes) has to destabilise the equilibrium and lead to the prominence instability and consequently to the CME eruption.

Most acceptable triggering mechanism for the prominence instability is connected to twisted magnetic configurations. It has been observed that the kink instability of magnetic flux ropes (Williams et al. \cite{Williams2005}), the torus instability (Filippov \cite{Filippov2013}, Zuccarello et al. \cite{Zuccarello2014}) or both together (Vasantharaju et al. \cite{Vasantharaju2019}) can lead to CME initiations.


Another phenomenon in the solar atmosphere is a coronal rain, cool and dense material condensing at solar coronal loops falling along its legs. The condensations are probably caused by thermal instability (Parker \cite{Parker1953}, Field \cite{Field1965}). Another type of coronal rain is related to solar prominences, where cool blobs detach from the prominence main body and fall down toward the photosphere. Recent SDO/AIA observations show the formation of the prominence by the condensation after CME, where most of the mass drained down through vertical downward flows (Liu et al. \cite{Liu2012}). Liu et al. (\cite{Liu2012}) concluded that these flows show up as cool and dense plasma blobs started falling at a height between 20-40 Mm during 30 min. Measurement of velocity, above the surface, has a narrow Gaussian distribution with values 30 km s$^{-1}$, while the derived accelerations have an average value of 46 m s$^{-2}$. It was assumed, that the thermal instability is responsible for the plasma condensation and hence the coronal rain through catastrophic cooling i. e. when the energy balance is violated as the radiative losses locally overcome to the heating input (Parker \cite{Parker1953}, Field \cite{Field1965}, Antiochos et al. \cite{Antiochos1999}, Schrijver \cite{Schrijver2001}, Vashalomidze et al. \cite{Vashalomidze2015}). Numerical simulations also show that the thermal instability is the reason for the formation of cold condensation and coronal rain (M\"uller et al \cite{Muller2003}, \cite{Muller2004}, \cite{Muller2005}).

\begin{figure}[t]
\vspace*{1mm}
\begin{center}
\includegraphics[width=\columnwidth]{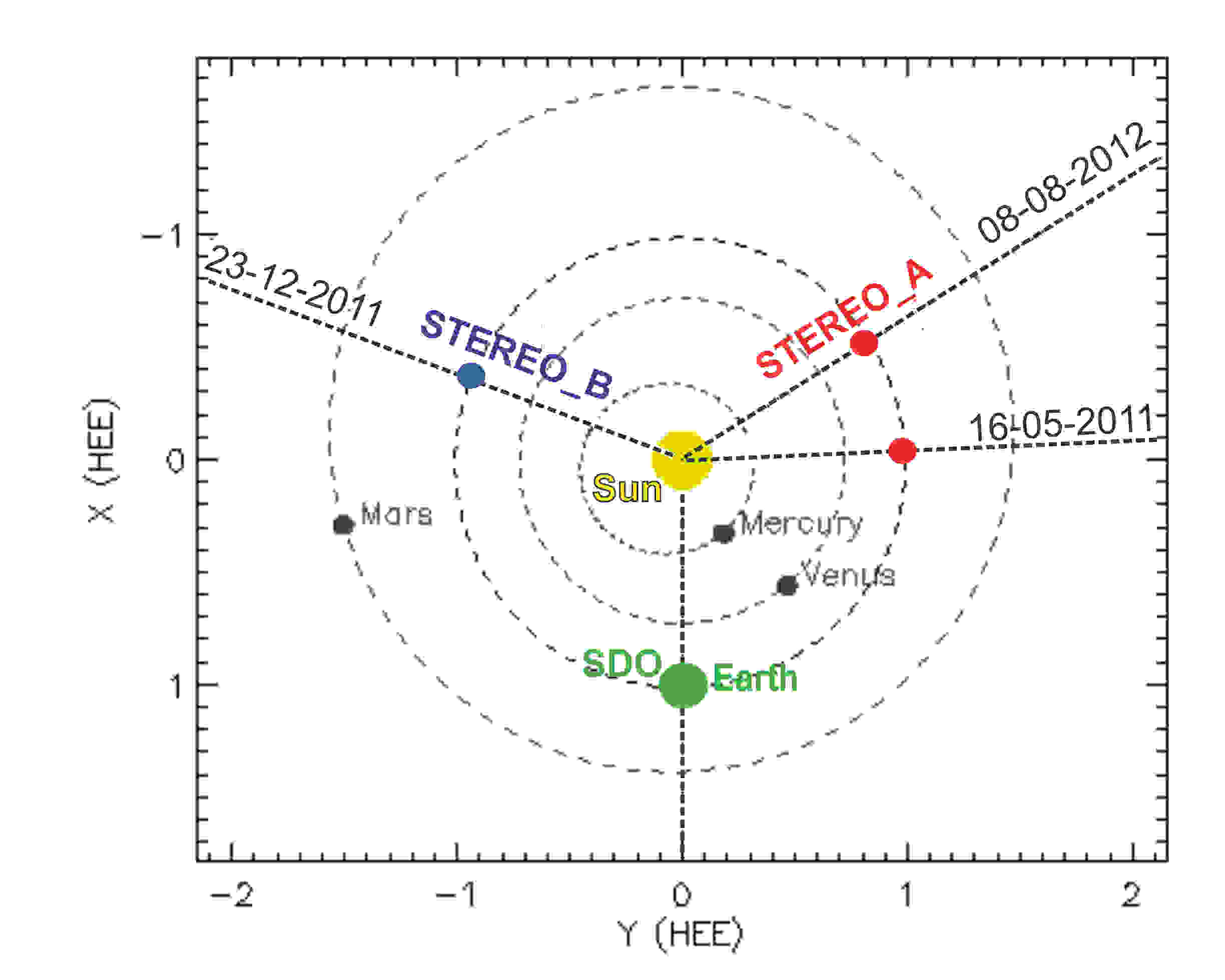}
\end{center}
\caption{Positions of SDO, Stereo\_A, and Stereo\_B in space during the observed events (the locations of spacecrafts and separation angles were obtained from STEREO web page: stereo-ssc.nascom.nasa.gov).}
\end{figure}

Equilibrium state of prominences is achieved owing to the balance between gravity and Lorentz forces (in the low-beta plasma). Massive plasma downflows from prominences in the form of coronal rain will lead to the decrease of the prominence mass, which may affect the equilibrium. Lorentz force may succeed over gravity after some time and the prominence may start to rise up slowly. This slow rising process may trigger the magnetic reconnection and/or instability, which can lead to CMEs. In this paper we present several observational evidences of prominence instability as triggered by massive downflows in form of coronal rain.

\section{Observation and Data Analysis}

We use observations from AIA on the board of SDO (Pesnell et al. \cite{Pesnell2012}, Lemen et al \cite{Lemen2012}) and Sun-Earth Connection Coronal and Heliosphere Investigation (SECCHI) on board of Solar Terrestrial Relations Observatory (STEREO).  SECCHI/EUVI (Extreme Ultraviolet Imager) takes images in 304 \AA, 171 \AA, 195 \AA, and 284 \AA\ channels with the spatial resolution of $1.6{"}$ per pixel of entire solar disk ($2048\times2048$ pixel images). We use only 304 \AA\ and 195 \AA\ data from STEREO/EUVI. AIA provides high spatial resolution images of $0.6"$ per pixel with a cadence of 12 seconds in multiple wavelength channels. We use three extreme ultraviolet (EUV) narrow bands at 304 \AA, 171 \AA\ and 193 \AA\ lines which correspond to the formation temperatures of $10^{4.7}\ K$, $10^{5.8}\ K$, and $10^{6.2}\ K$, respectively.

\begin{figure}
\begin{center}
\includegraphics[width=9cm]{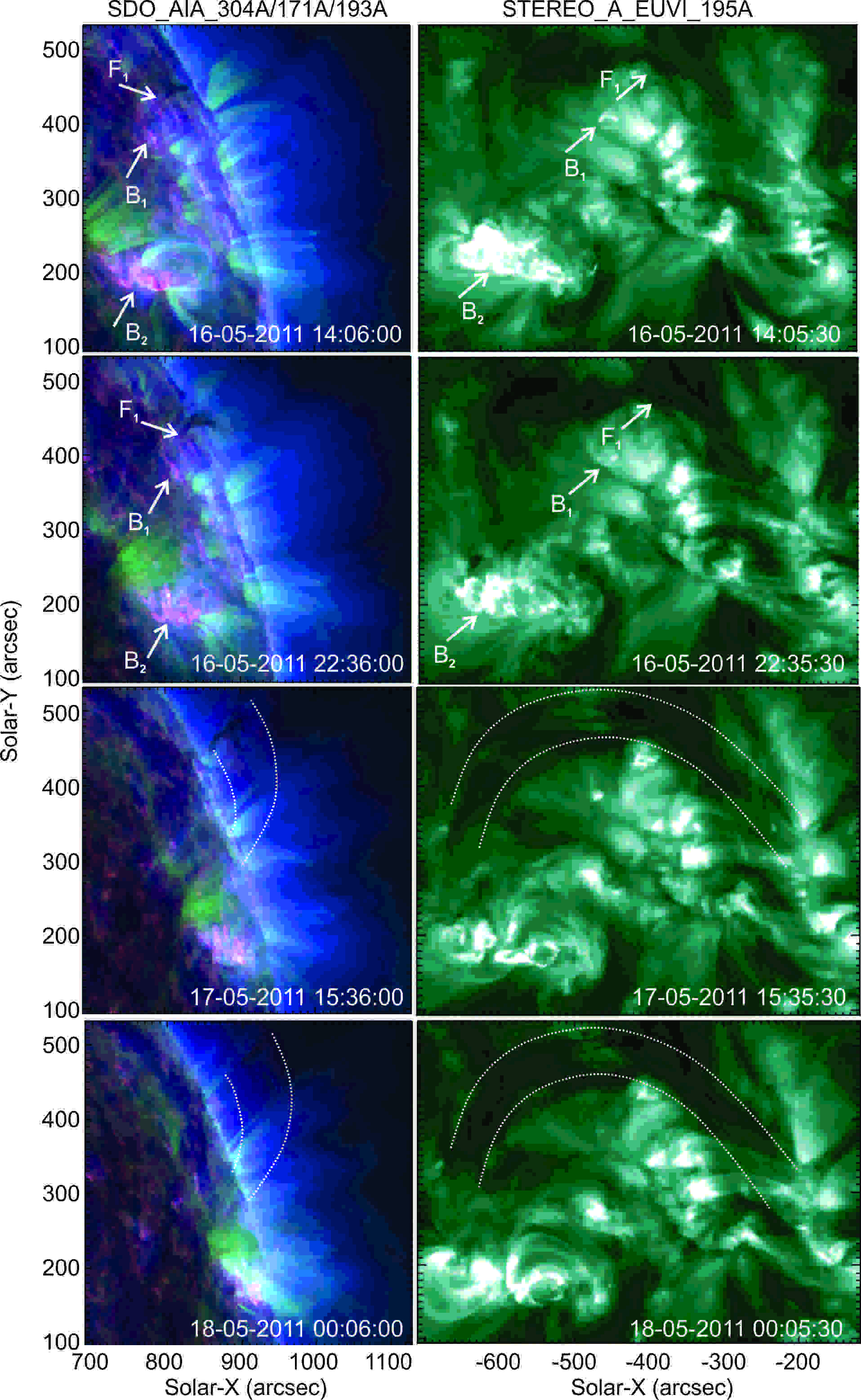}
\end{center}
\caption{Evolution of the Prominence/filament from 14:06 UT 16 May to 00:06 UT 18 May, 2011 in SDO/AIA and Stereo\_ A/EUVI. Left panels show composite images of 304 \AA, 171 \AA, and 193 \AA\ wavelengths from SDO/AIA. Right panels show 195 \AA\ wavelength of Stereo\_ A/EUVI. The white arrows in the top four frames indicate the specific points on images from both spacecraft. $F_{1}$ points correspond to the footpoint of tornado-like structure. The brightening is marked with $B_{1}$ and $B_{2}$ points for identifying areas of interest. Approximate boundaries of prominence are shown by white doted-curved lines on lower four panels.}
\end{figure}

\begin{figure*}
\begin{center}
\includegraphics[width=16cm]{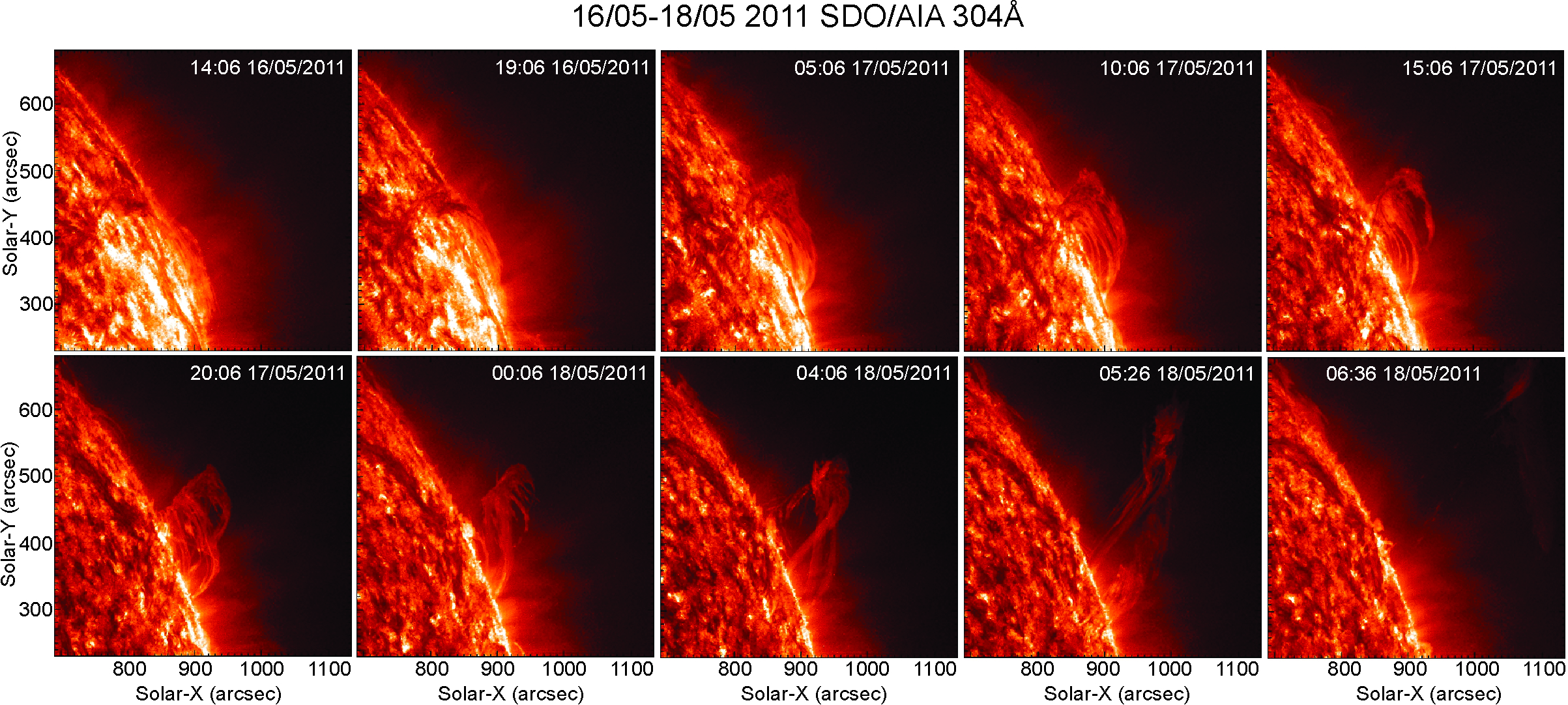}
\end{center}
\caption{The evolution of the prominence in 304 \AA\ line of SDO/AIA  during 14:06 UT 16 May and 06:36 UT 18 May, 2011.}
\end{figure*}

\begin{figure*}
\begin{center}
\includegraphics[width=15cm]{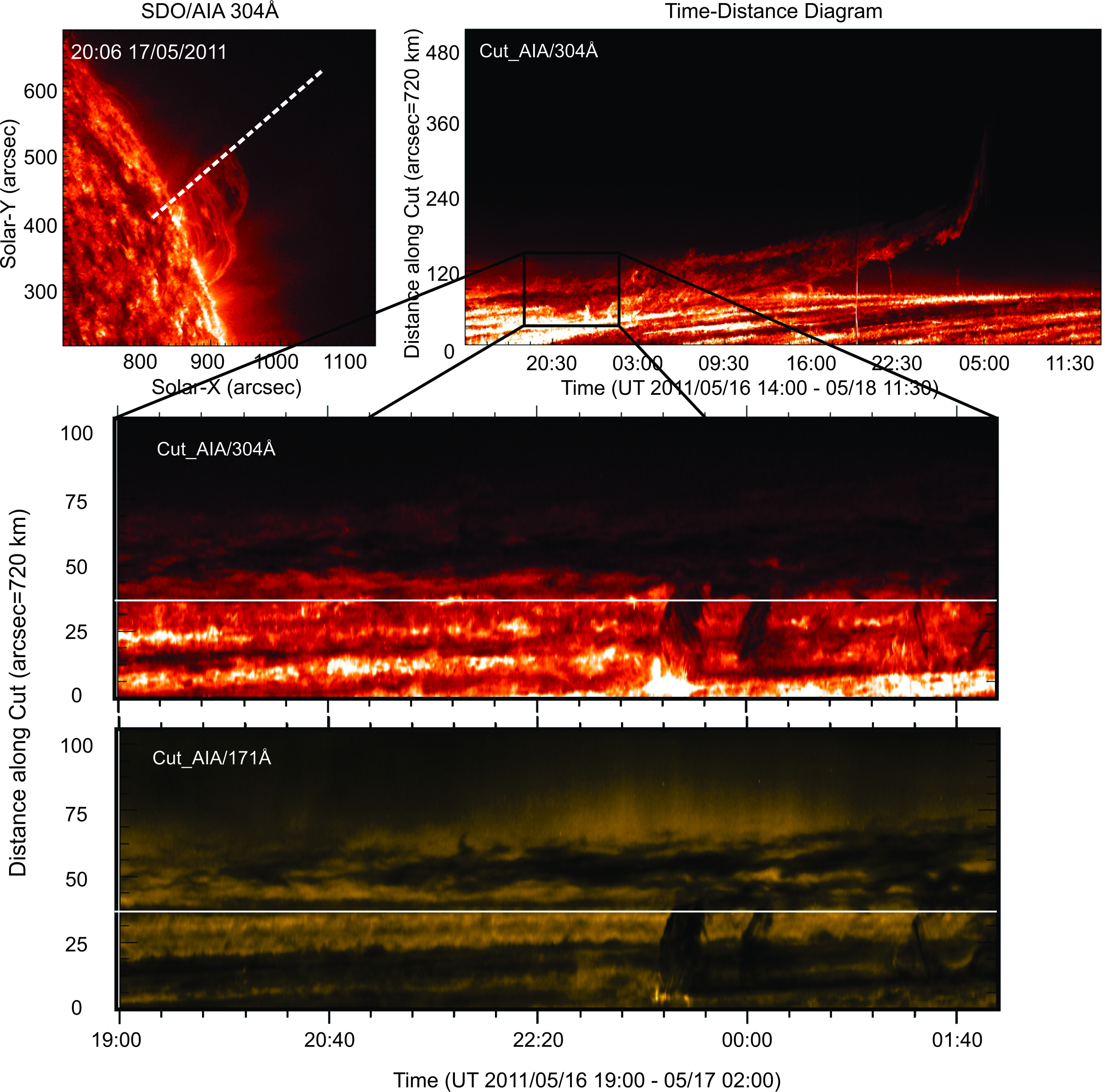}
\end{center}
\caption{Space-time diagram of prominence rise at a fixed cut in 304 \AA\ line of SDO/AIA during 14:06 UT 16 May and 11:30 UT 18 May, 2011 (top right panel). The location of the cut is shown by white dashed line on top left panel. The middle and the lower panels show zoomed black box corresponding to the interval of 19:00 UT 16 May - 02:00 UT 17 May (i. e. before the start of slow rising) in  304 \AA\ (middle panel) and 171 \AA\ (lower panel) lines.}
\end{figure*}

We analysed three different events observed during the years of 2011-2012: on May 16-18, 2011, December 22-24, 2011 and August 07-08, 2012. Locations of STEREO and SDO spacecrafts in space allowed us to observe each event from different angles, hence to follow the structure and dynamics of selected prominences in detail. Positions of SDO, Stereo\_A and Stereo\_B in space during the three events are shown on Fig. 1. The first (May 16-18, 2011) and third (August 07-08, 2012) events were observed by SDO and Stereo\_A with the the separation angle between the spacecrafts of $+93^{\circ}$ and $+122^{\circ}$, respectively. The second event of December 22-24, 2011 was observed by SDO and Stereo\_B with the separation angle of $-110^{\circ}$.



In order to identify prominences on both SDO and STEREO datasets, we used the SSW (Solar Software) STEREO procedure wcs\_convert\_diff\_rot, which takes heliographic coordinates of set point on image from one spacecraft and applies a differential rotation model (another stereo procedure diff\_rot.pro) to match the position of the point on another spacecraft image.

We use the path-draw functionality of CRISPEX (The CRISP Spectral Explorer) (Vissers et al. \cite{Vissers2012}), for tracing falling material along visible trajectory and space-time diagrams along traced paths.
From  the space-time diagrams, we determined projected velocities and accelerations by using the CRISPEX auxiliary program TANAT (Vissers et al. \cite{Vissers2012}).

\section{Results}

\subsection{The event of May 16-18, 2011}

The separation angle between SDO and Stereo\_A was $+93^{\circ}$ on May 16, 2011, therefore the eastern edge of the SDO image (the eastern limb) lies on $-3^{\circ}$ longitude on Stereo\_ A image.

The target prominence first appeared at the eastern limb on SDO images at 05:00 UT on May 16, 2011. At the same time, it was seen on disk near the western limb in Stereo\_ A. Figure 2 shows the evolution of the prominence from SDO and Stereo\_A during 14:06 UT 16 May and 00:06 UT 18 May, 2011. The left panels show composite images from SDO/AIA and the right panels show 195 \AA\ wavelength of Stereo\_ A/EUVI. White arrows show the points, which were identified with both spacecraft observations.

A tornado-like structure started to rise up at  05:00 UT on May 16 (indicated by $F_{1}$ on Fig. 2). To identify the area of interest more precisely, we also marked two small brightening on images of both spacecrafts (indicated by $B_{1}$ and $B_{2}$). White doted-curved lines on the lower panels on Figure 2 show the mean edges of prominence tube.

Figure 3 shows the evolution of the prominence in 304 \AA\ line of SDO/AIA during 14:06 UT (May 16)-06:36 UT (May 18). Massive fall of coronal rain from the prominence main body started on 22:00 UT (May 16) and continued until 02:00 UT (May 18). After almost 28 hrs of coronal rain, the prominence started to be destabilised and finally erupted as a CME on May 18, 2011.


Figure 4 shows a process of prominence rise in detailed using a vertical space-time cut at particular location indicated by a white dotted line on upper left panel. Upper right panel shows the space-time diagram during the whole interval of the prominence evolution. it is seen that the whole process of rising may be formally divided into two phases: slow rise phase (from about 03:00 UT, May 17 to 04:00 UT, May 18)  and fast rise phase (from 04:00 UT, May 18). The fast rise phase corresponds to the final destabilisation and eruption of the prominence as a CME. The slow rise phase is probably connected to the slow loosing of mass due to the rain (see subsection 3.4).

In order to study the coronal rain, we traced the falling blobs along visible trajectory from the SDO images and have drawn their paths using the path-draw functionality of CRISPEX  (Vissers et al. \cite{Vissers2012}). We identified 12 visible trajectories of falling plasma, which are shown as white dotted lines on upper left panel of Fig. 5. Then we constructed space-time diagrams for each trajectory. Lower panel of Fig. 5 shows the space-time diagram along the curved trajectory of number 6 indicated by the white arrow on the upper left panel. The white dashed curves on this panel show well-defined trajectories of falling blobs. We fitted trajectories of all cuts with a polynomials and determined the average velocity of falling material to $\overline{v}=$ 23.5 km s$^{-1}$. Acceleration of coronal rain blobs is estimated to be smaller than the free fall in the solar atmosphere as it is typical for the rain.

\begin{figure*}
\begin{center}
\includegraphics[width=14cm]{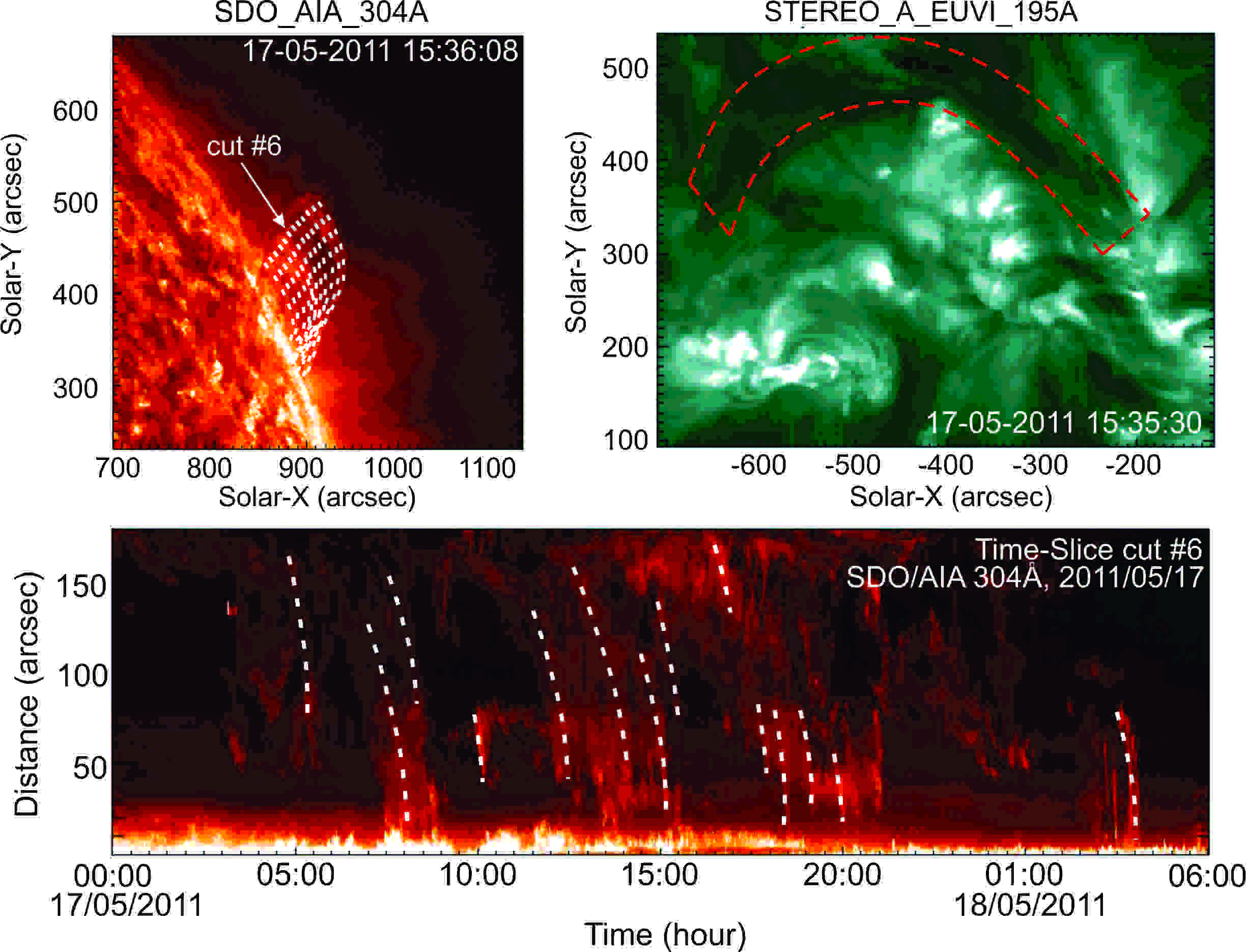}
\end{center}
\caption{Simultaneous Images taken at 15:36 UT on May 17, 2011, from SDO/AIA and Stereo\_ A/EUVI. Top left panel show composite image of 12 visible trajectories of falling plasma with white dotted lines, from SDO/AIA. The top right frame shows the area of prominence tube with a red dashed line in 195 \AA\ wavelength from Stereo\_ A/EUVI spacecraft. The bottom image shows the space-time diagram of falling plasma obtained from a typical cut along the traced path (the white arrow indicate the starting point and the position of the cut 6 trajectory on the top right panel).}
\end{figure*}

Prominence equilibrium is achieved by balance between gravity and the Lorentz force of prominence magnetic field. Coronal rain obviously took away a part of the prominence mass.  Therefore, the reduction of the prominence mass may lead to the violation of the equilibrium and hence to the observed slow rise of the prominence. One can roughly estimate the mass loss of the prominence after 28 hours of coronal rain. The total width of the 12 threads, where the coronal rain blobs were falling down, is around 12 SDO/AIA pixels, which corresponds to 0.864 $\times$ 10$^{9}$ cm  (SDO/AIA pixel equal 720 km). Assuming the width as the tube diameter gives the total cross section of coronal rain paths as 0.59 $\times$ 10$^{18}$ cm$^{2}$. Using the typical electron number density in prominence cores as $n_{e}=10^{10} $ (Labrose et. al \cite{Labrose2010}), one can estimate prominence mean mass density as $\rho$=1.67 $\times$10$^{-14}$ g cm$^{-3}$. Then the mean speed of $\overline{v}=$ 23.5 km s$^{-1}$ lead to the total mass flux of 2.32 $\times$ 10$^{9}$ g s$^{-1}$. Therefore, the estimated mass loss after 28 hr coronal rain is $\approx$ 2.34 $\times$ 10$^{15}$ g.

\begin{figure}
\begin{center}
\includegraphics[width=6cm]{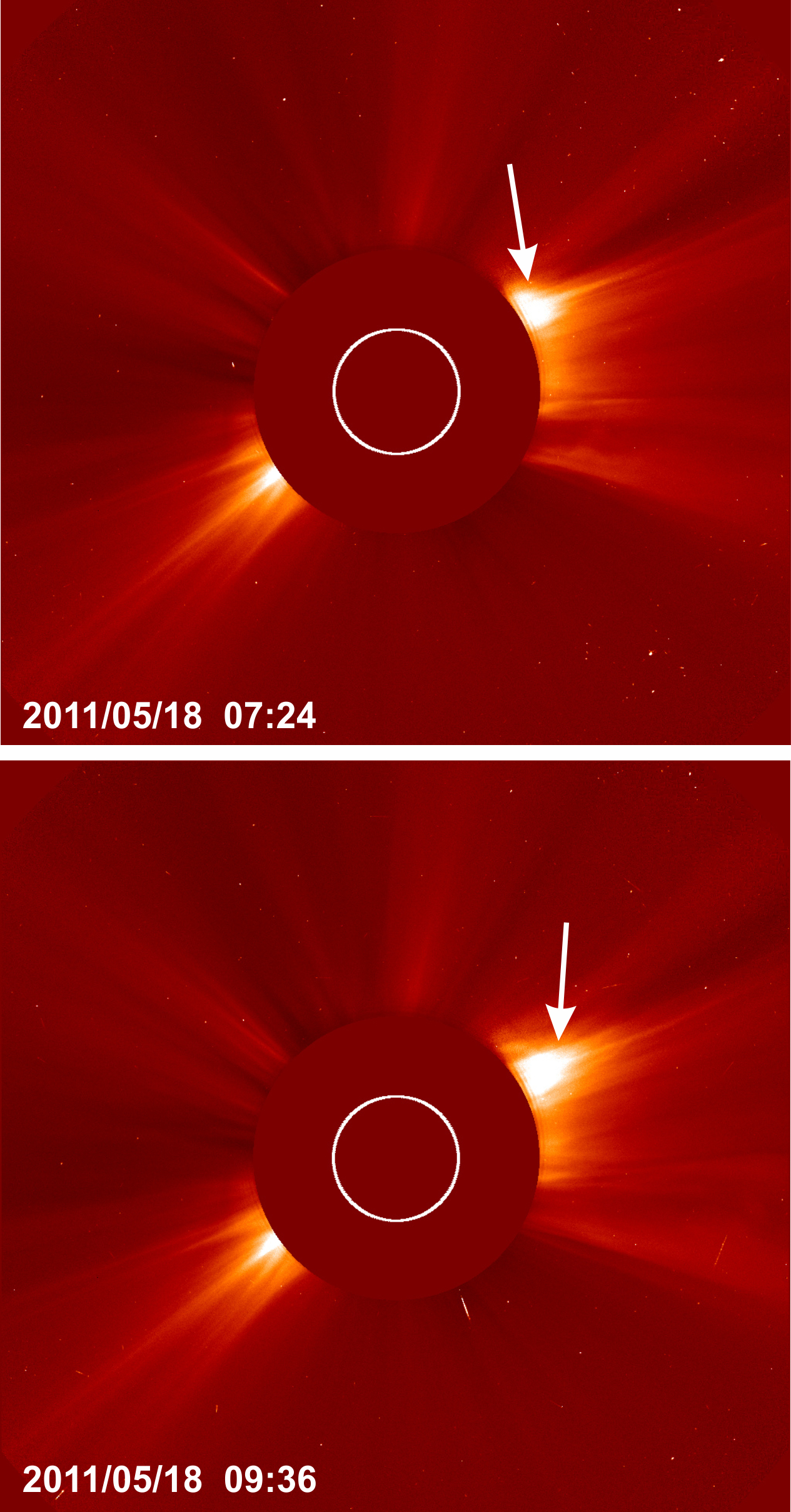}
\end{center}
\caption{CME eruption after the prominence instability as seen from LASCO/C2/SOHO. C2 displays an image in white light at distance 2.1-6 solar radii. The white arrows indicate the location of the CME. The two images show the time evolution of the CME at 07:24-09:39 UT on May 18, 2011.}
\end{figure}

Simultaneous observations from  SDO and STEREO allow us roughly estimate the prominence volume. Prominence length was estimated from Stereo data  as 500 px, which with the STEREO pixel resolution of 1.6 arc sec gives 5.76 $\times10^{10}$ cm (see red dashed area on right upper panel of Fig. 5). The mean halfwidth of the prominence can be estimated as 30 px leading to 2.16 $\times 10^{9}$ cm (Fig. 5 upper right panel). Assuming the halfwidth as the mean radius of cylindrical volume of the prominence, one can calculate the cross section as 1.47$\times$10$^{19}$ cm$^{2}$ and hence  the prominence volume as 8.4 $\times$ 10$^{29}$  cm$^{3}$. Then the prominence total mass is roughly estimated as 1.4 $\times$ 10$^{16}$ g. The estimated mass loss suggests that around 17 \% of the prominence mass was taken away by the coronal rain before the final instability set up at 04:00 UT, May 18. We note however that this estimation corresponds to the coronal rain only along one leg of the prominence as we could not follow to dynamics of the another leg. If we suppose the similar mass flux along another leg, then
the total mass loss can be estimated as 34 \% of the total mass.

\begin{figure}
\begin{center}
\includegraphics[width=9cm]{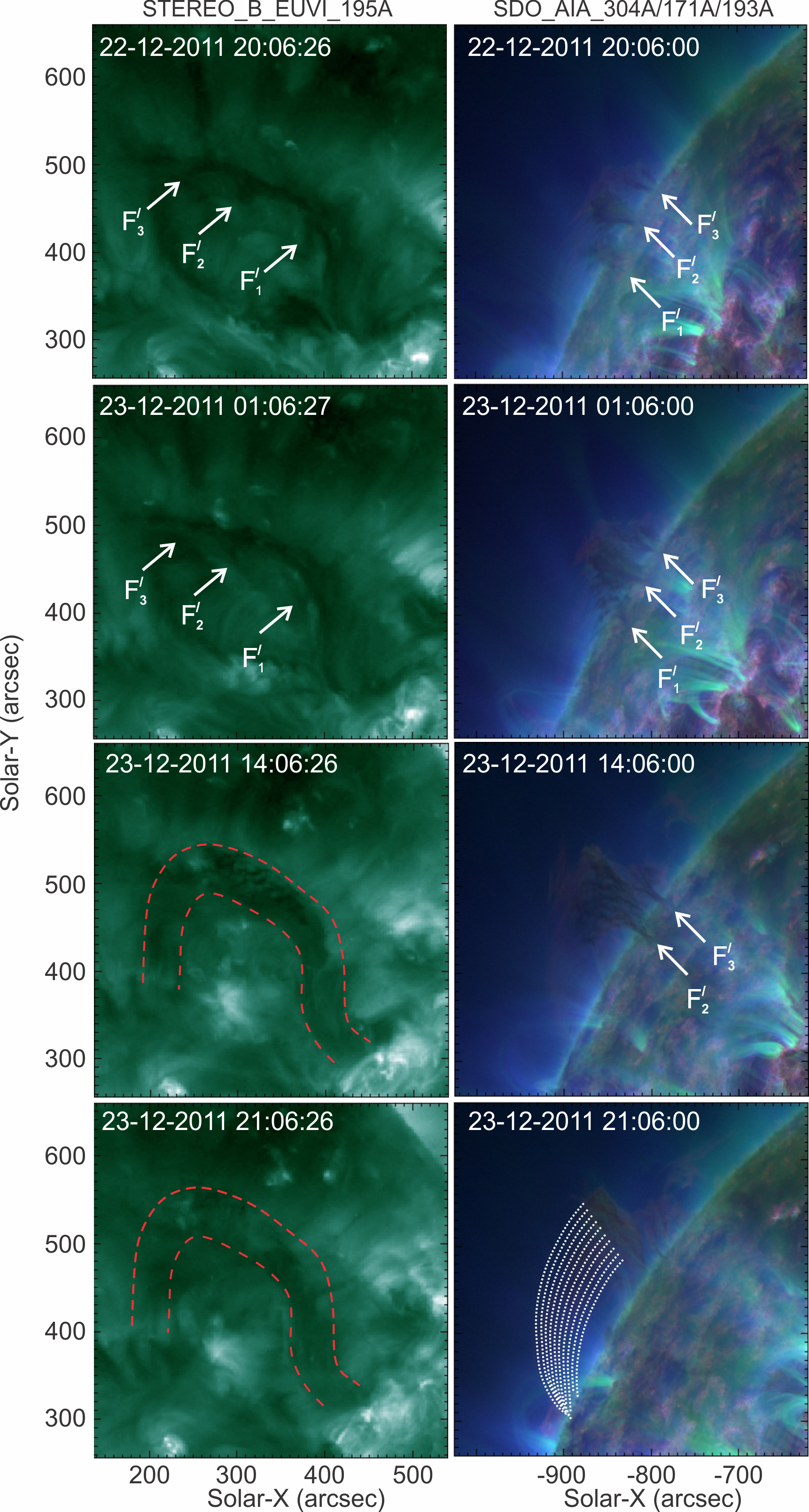}
\end{center}
\caption{Evolution of target prominence/filament from 20:06 UT 22 December until 21:06 UT 23 December, 2011, in SDO and Stereo\_B spectral lines. The right column shows composite images of 304 \AA\, 171 \AA\, and 195 \AA\  lines from SDO/AIA and the left frames show 195 \AA\ line of Stereo\_ B/EUVI. White arrows marked with ${F_{1}}^{'}$, ${F_{2}}^{'}$ and ${F_{3}}^{'}$ show the locations of prominence footpoints in both spacecrafts. Red dashed curved lines show approximate boundaries of prominence, while the white doted curved lines on right lower panel show the trajectories of coronal rain.}
\end{figure}

When the prominence disappeared in 304 \AA\ line after the eruption, it was evidently detected in white light images of SOHO (Solar and Heliospheric Observatory) instrument LASCO (The Large Angle and Spectrometric Coronagraph) as a CME. Figure 6 shows the dynamics of the CME in the outer corona as observed from LASCO/C2, which covers a distance of 2.1-6 solar radii. According to the LASCO CME catalog (Yashiro et al. \cite{Yashiro2004}), the first appearance of the CME in the field of view C2 coronagraph is reported at 07:00:05 UT, May 18, 2011 with a speed of 141 km s$^{-1}$. The CME appeared with an angular width of $36^{\circ}$ in the northeast limb (see Figure 6).

\subsection{The event of December 22-24, 2011}

The separation angle between Stereo\_B and SDO was $-110^{\circ}$ on December 22, 2011, hence the western limb of the SDO image corresponds to the $+20^{\circ}$ longitude on Stereo\_ B image.

The target prominence first appeared at the western limb on SDO spacecraft at 11:00 UT on December 22, 2011. At the same time, it was seen on disk near the eastern limb in Stereo\_ B. Figure 7 shows images from 20:06 UT 22 December until 21:06 UT 23 December 2011. The left panels show 195 \AA\ wavelength of Stereo\_ B/EUVI and the right panels show composite images from SDO/AIA.

With the red dashed-curved line on the lower panels on Figure 7, we marked approximate boundaries of prominence tube. We identified 10 visible trajectories of falling plasma, which are shown as white dotted lines on lower right panel of Fig. 7.

Figure 8 shows the evolution of the prominence in 304 \AA\ line of SDO/AIA during 20:06 UT (December 22)-00:46 UT (December 24). Massive coronal rain started to flow down from the prominence main body on 20:00 UT (December 22) and continued until 21:45 UT (December 23) i. e. almost 26 hours. The duration of coronal rain is similar to the previous case. After the coronal rain the prominence started to be unstable on December 23 and finally erupted on December 24, 2011.

As in the previous case, we fitted trajectories with a polynomials and determined the average velocity of coronal rain as $\overline{v}=$ 52.6 km s$^{-1}$. Acceleration of coronal rain blobs is again smaller than the free fall. The total width of the 10 threads is around 10 pixels and consequently 7.2 $\times$ 10$^{8}$ cm  (SDO/AIA pixel equal 720 km) with the total cross section of coronal rain paths of 4.08 $\times$ 10$^{17}$ cm$^{2}$. This gives the total mass flux of 3.58 $\times$ 10$^{10}$ g s$^{-1}$ along one leg of the prominence leading to the mass loss of 3.35 $\times$ 10$^{15}$ g in 26 hours.

Prominence mean length and halfwidth are estimated from Stereo\_ A/EUVI as $\sim$ 4 $\times10^{10}$ cm  (see red dashed area on left below panel of Fig. 7) and $\sim$ 5.76 $\times 10^{9}$ cm (Fig. 7 lower left panel), respectively, which lead to the prominence volume as $\sim$ 1.05 $\times$ 10$^{30}$  cm$^{3}$. Using the mean density of the prominence (see previous subsection) one may estimate the prominence mass as 1.75 $\times$ 10$^{16}$ g. Hence due to the coronal rain the prominence lost around 19 \% of its mass before it became unstable. Again if we consider the symmetric mass flux along the both legs, then the mass loss rise up to 38 \%.

After the eruption, the prominence appeared in field of view C2 coronagraph at 00:36:05 UT, December 24 as a CME with a speed of 475 km s$^{-1}$. Figure 9 shows the dynamics of the CME in the outer corona as observed from LASCO/C2. The CME appeared with an angular width of $61^{\circ}$ in the northwest limb. Blue lines on the left panels of the time difference images shows the position of the leading edge of the CME in the outer corona, while the red lines show an approximate outline to the leading edge that was created by using a segmentation technique (Olmed at el. \cite{Olmed2008}), at the distance of 6.1 solar radii (images are taken from George Mason University Space Weather Lab project SEEDS (Solar Eruptive Events Detection System))(Fig.9).

\begin{figure*}
\begin{center}
\includegraphics[width=16cm]{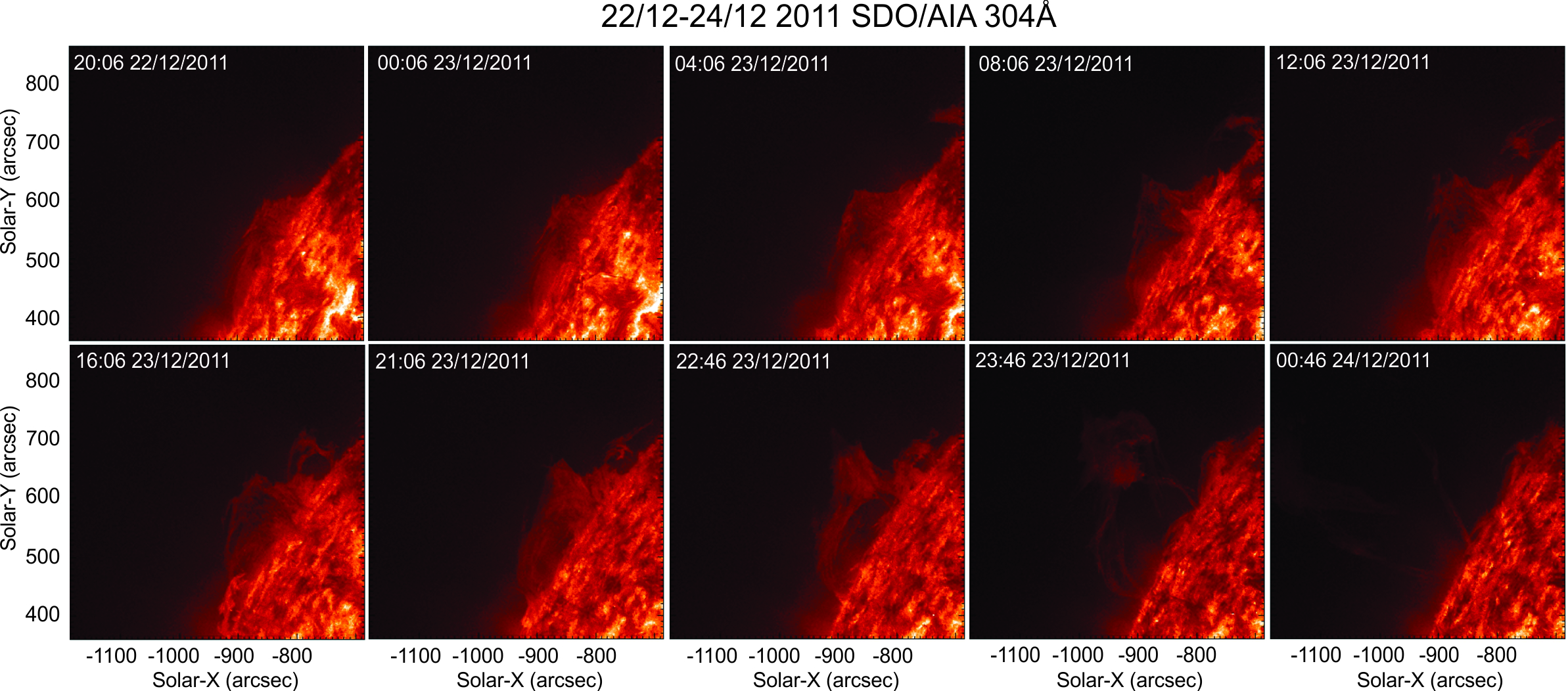}
\end{center}
\caption{The evolution of observed prominence/filament with accompanied eruption in 304 \AA\ wavelength of SDO/AIA from 20:06 UT 22 December until 00:46 UT 24 December 2011.}
\end{figure*}

\subsection{The event of August 07-08, 2012}

The separation angle between SDO and Stereo\_ A was $-122^{\circ}$ on August 07, 2012, therefore the eastern limb of the SDO images lie on $-32^{\circ}$ latitude on Stereo\_ A images.

The target prominence first appeared at the eastern limb on Stereo\_ A  at 09:06 UT on August 07, 2012. It was seen on disk near the western limb in SDO. Figure 10 shows the prominence from 09:06 UT 07 August until 02:06 UT 08 August 2012. The white dashed-curved line on the lower left panels on Figure 10 shows approximate boundaries of prominence tube. We identified 10 visible trajectories of falling plasma, which are shown as white dotted lines on lower right panel of Fig. 10.

\begin{figure}
\begin{center}
\includegraphics[width=8cm]{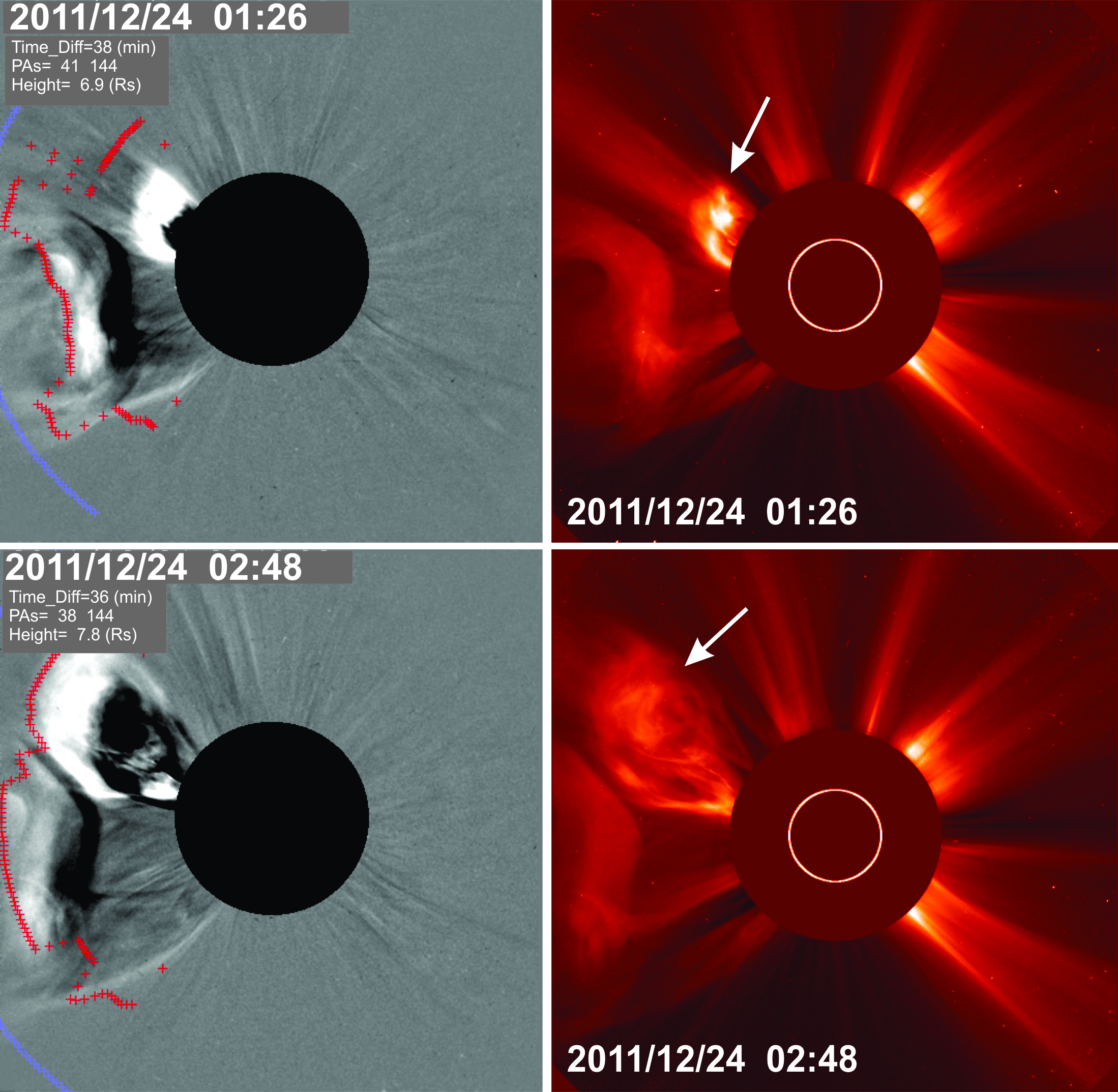}
\end{center}
\caption{CME eruption after the prominence instability as seen from LASCO/C2/SOHO. Left panels show the running difference images, where the time difference is $\sim$36 min. The blue line indicates to the position of the leading edge, while the red line indicates an approximate outline to the leading edge. Right panels show images from LASCO/C2 coronagraph. The white arrows indicate the location of the CME. The two images show the time evolution of the CME at 01:26-02:48 UT on December 24, 2011.}
\end{figure}

Figure 11 shows the evolution of the prominence in 304 \AA\ line of Stereo\_ A/EUVI during 09:06 UT (August 07)-04:06 UT (August 08). Massive fall of coronal rain started from the prominence main body on 09:06 UT (August 07) and continued until 04:06 UT (August 08) i. e. almost 18 hours. The duration of coronal rain is less than in the previous cases. After the coronal rain the prominence started to be destabilised on August 07 and finally erupted as a CME on August 08, 2011.

The average velocity of coronal rain was estimated as $\overline{v}=$ 64 km s$^{-1}$. The total width of the 10 threads is around 1.15 $\times$ 10$^{9}$ cm with the total cross section of coronal rain paths as 1.04 $\times$ 10$^{18}$ cm$^{2}$. This gives the total mass flux of 1.11 $\times$ 10$^{10}$ g s$^{-1}$ leading to the mass loss of 7.22 $\times$ 10$^{15}$ g in 18 hours. Prominence mean length and halfwidth are estimated from SDO/AIA as $\sim$ 1.01 $\times10^{11}$ cm  (see white dashed area on left below panel of Fig. 10) and $\sim$ 2.88 $\times 10^{9}$ cm (Fig. 10 lower left panel), respectively, which lead to the prominence mass as 4.39 $\times$ 10$^{16}$ g. Then the mass loss due to the coronal rain along one leg led to the reduction of the prominence mass with 17 \%.

Figure 12 shows the time difference images of the CME in the outer corona as observed from LASCO/C2. The CME appeared in the field of view C2 coronagraph at 05:48:07 UT, August 08, 2012 with a speed of 355 km s$^{-1}$. On Figure 12, Blue lines on the left panels of the time difference images shows the position of the leading edge of the CME in the outer corona, while the red lines show an approximate outline to the leading edge that was created by using a segmentation technique (images are taken from George Mason University Space Weather Lab project SEEDS (Solar Eruptive Events Detection System)).

\begin{figure}
\begin{center}
\includegraphics[width=8cm]{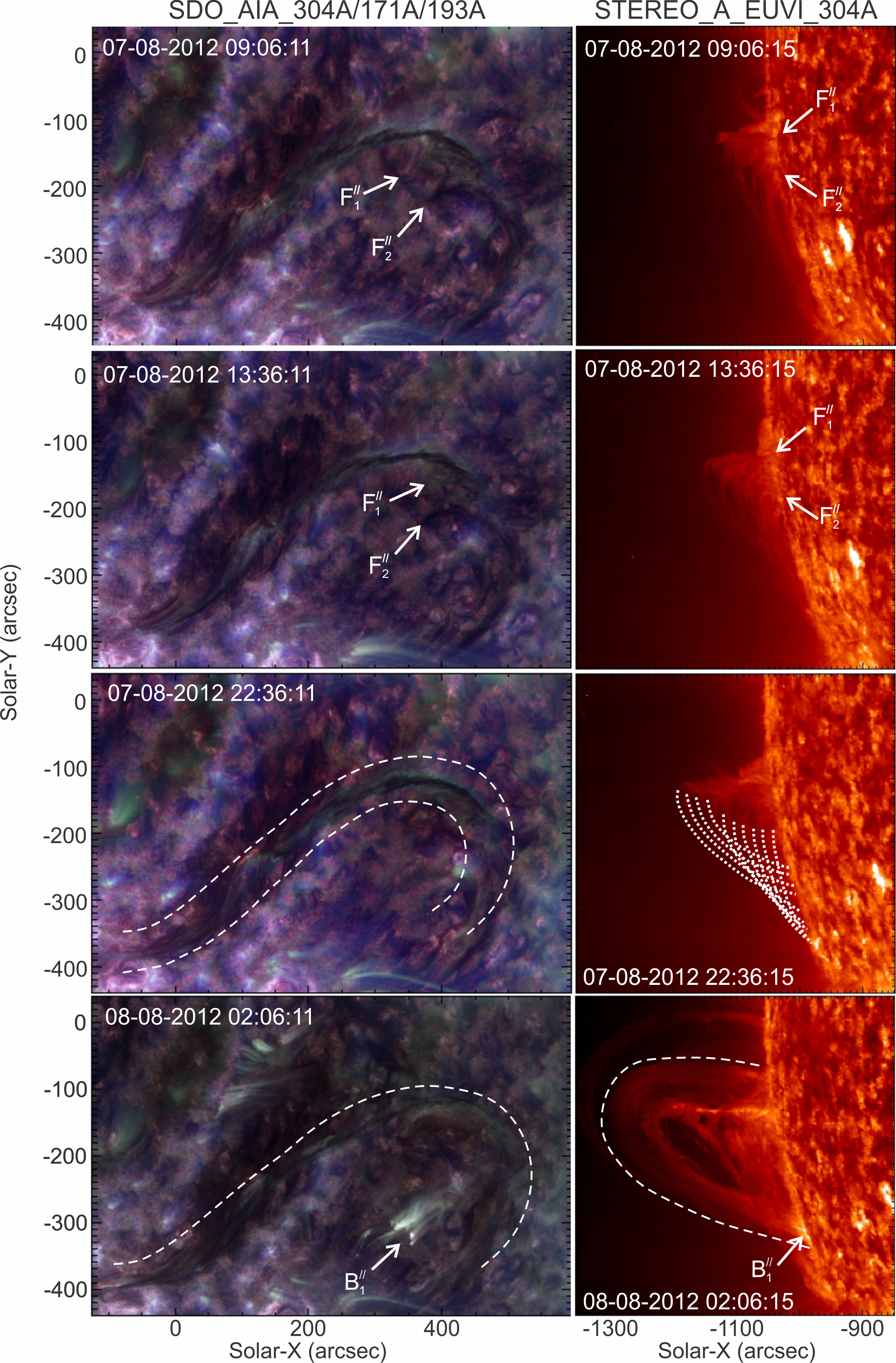}
\end{center}
\caption{Evolution of the Prominence/filament from 09:06 UT 07 August and 02:06 UT 08 August, 2012 in SDO/AIA and Stereo\_ A/EUVI. Left panels shows composite images of 304 \AA\, 171 \AA\, and 195 \AA\ wavelength from SDO/AIA. Right panels show 304 \AA\ wavelength of Stereo\_ A/EUVI. The white arrows in the top four frames, marked with ${F_{1}}^{''}$ and ${F_{2}}^{''}$ show the locations of prominence footpoints in both spacecrafts. White dashed curved lines indicate approximate boundaries of prominence, while the white doted curved lines on left lower panel show the trajectory of coronal rain. The brightening is marked with ${B_{1}}^{''}$ point for identifying areas of interest.}
\end{figure}

\subsection{Possible mechanisms for the initial rise of prominence}

As we already noted, a prominence is the result of equilibrium between magnetic and gravitational forces in the coronal low plasma-beta regime. Violation of the equilibrium resulting the initial slow rise of prominence could be caused by two reasons. First, the thermal instability may result in the fall of coronal rain blobs, which reduces the prominence mass and hence leads to the slow rise of prominence. Second, the violation of equilibrium may be caused by a magnetic process (e. g. instability), which leads to the slow rise of prominence  and consequently to the fall of prominence mass in form of blobs as observed. In both cases, the coronal rain blobs are important ingredients in the process of slow rising. But it is of vital importance to understand the triggering mechanism for initial rise of prominence body.

In order to study the process of initial rise we considered the event of May 16-18, 2011 in detailed.  We zoomed out the space-time diagram on Figure 4 around the start of coronal rain i.e. 22:00 UT (May 16). Two lower panels show the zoomed interval in two spectral lines. It is seen that the prominence was relatively stable until 00:30 UT (May 17) and then it started to rise up slowly. Hence the slow rise started after 2.5 hours of the intensive coronal rain process (to follow the process of coronal rain during 22:00 UT (May 11) -  00:30 UT (May 17), see accompanied movie). Obviously, when the prominence started to rise, the fall of plasma blobs from the prominence body is enhanced along magnetic field lines due to geometrical effects. But the initial rise of prominence seems to be caused by coronal rain, which presumably is the result of thermal instability.

Vasantharaju et al. (\cite{Vasantharaju2019}) suggested that the slow rise phase can be connected with the ideal kink instability of flux rope. Let us examine this possibility for the event of May 16-18, 2011. In this case, the slow rise process continued from 03:00 UT, May 17 to 04:00 UT, May 18 , i.e. about 25 hours (Figure 4). Growth time of ideal kink instability (Velli et al. \cite{Velli1990}, T\"or\"ok et al. \cite{Torok2004}, Zaqarashvili et al. \cite{Zaqarashvili2010}) is estimated as 10-100 transverse Alfv\'en time (the ratio of tube radius and Alfv\'en speed). The radius (or half width) of prominence flux tube is around $\sim$ 20 Mm in our case, while the expected Alfv\'en speed in prominences is $\sim$ 10-20 km/s. Then the growth time for ideal kink instability can be estimated as 0.3-0.5 hours, which is much shorter than the slow rise time of the targeted prominence (25 hours). Therefore, the kink instability is unlikely the reason for the slow rise of prominence body.

To check the dynamics of magnetic field configuration before and during the slow rise phase,  we performed magnetic field extrapolation (potential-field source-surface - PFSS) using the standard PFSS package 1 available under SolarSoftWare (SSW).  Figure 13 shows that  the magnetic field does not display any structural change during the slow rise phase, which indicates to the absence of magnetic instability in this interval (the extrapolated field lines before and during slow rise phase are shown on the upper and lower panels, respectively). This also indicates to the static configuration of photospheric magnetic field during the observed interval, which may rule out the role of photospheric changes in the slow rise process.

\begin{figure*}
\begin{center}
\includegraphics[width=16cm]{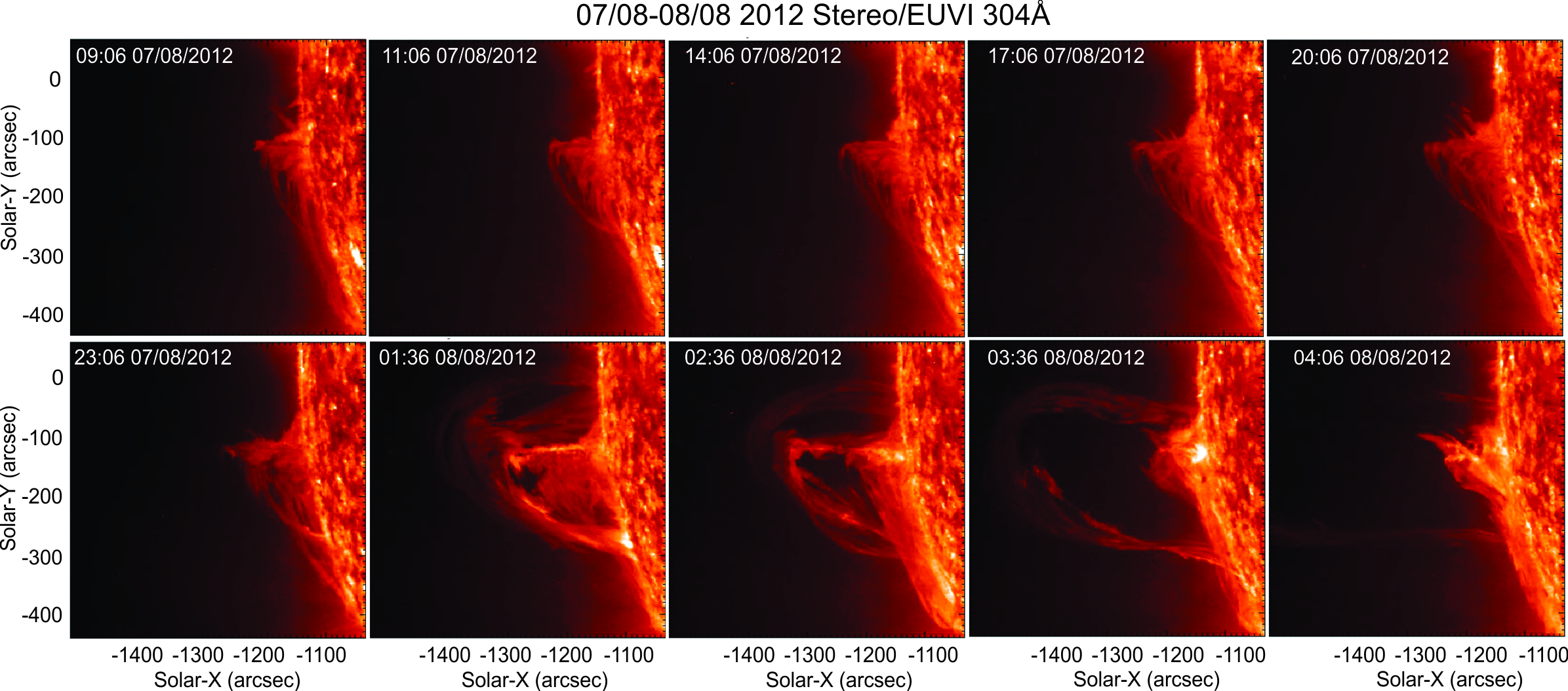}
\end{center}
\caption{Evolution of observed prominence in 304 \AA\ wavelength of STEREO/EUVI between 09:06 UT 07 August and 04:06 UT 08 August 2012.}
\end{figure*}

On the other hand, the fast rise phase probably corresponds to the magnetic instability, which can be processed either via the breakout model of coronal arcade (Antiochos et al. 1999) or torus instability of flux ropes (T\"or\"ok and Kliem 2007, Filippov \cite{Filippov2013}, Zuccarello et al. \cite{Zuccarello2014}). We can not make firm observational support which of the two models work in our case. Figure 3 clearly shows the flux rope structure of the prominence in the fast rise phase, which may support the torus instability, but the breakout model can not be completely ruled out. The prominence is erupted as a CME into the outer corona as a result of the instability.  Figure 14 shows the schematic picture of the whole process, which could be relevant to our case.

\begin{figure}
\begin{center}
\includegraphics[width=8cm]{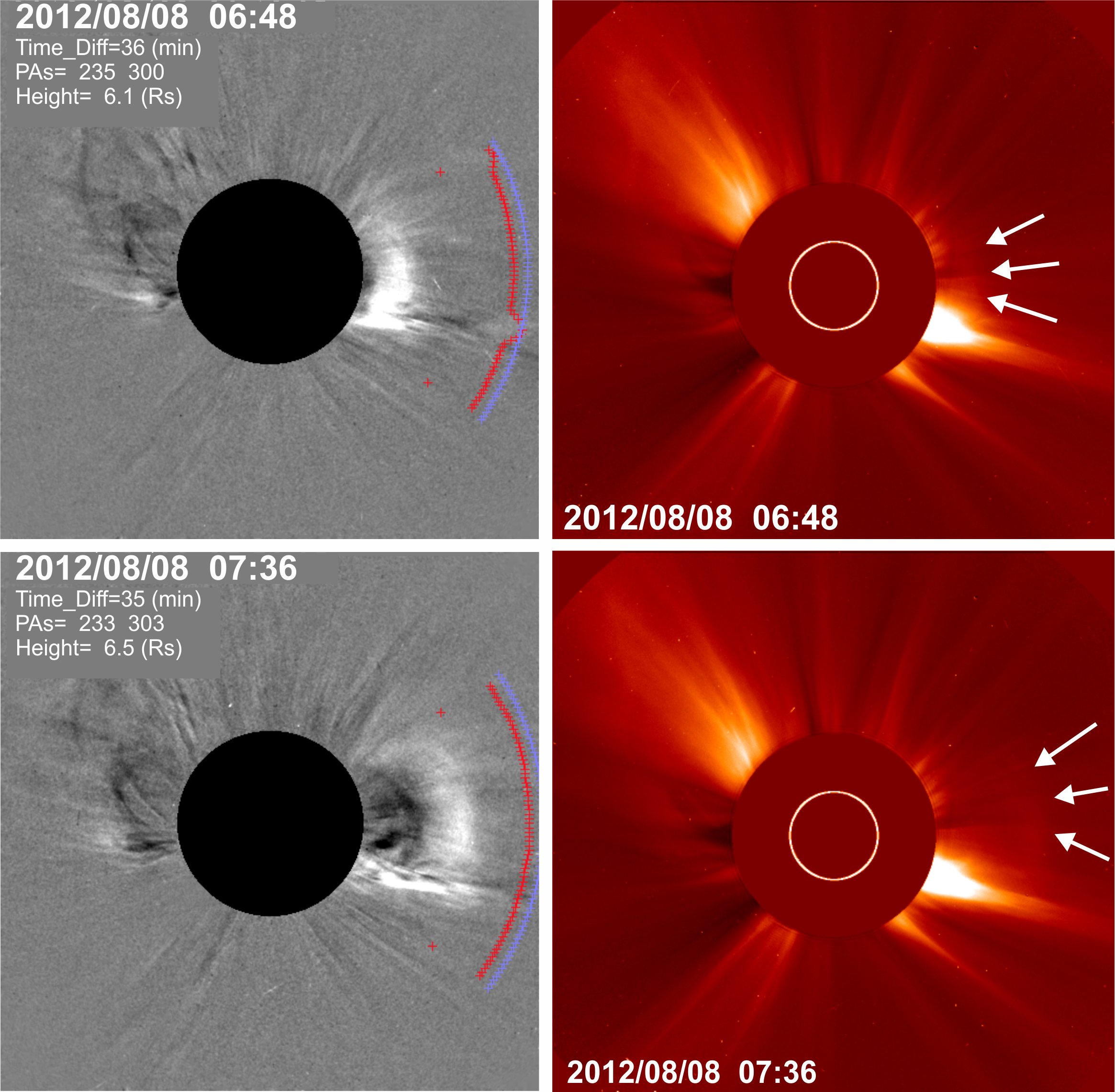}
\end{center}
\caption{CME eruption after the prominence instability as seen from LASCO/C2/SOHO. Left panels show the running difference images, where the time difference is $\sim$35 min. The blue line indicates to the position of the leading edge, while the red line indicates an approximate outline to the leading edge. The white arrows indicate the location of the CME. The two images show the time evolution of the CME at 06:48-07:36 UT on August 08, 2012.}
\end{figure}
{
During the slow rise phase, the prominence body evolves slowly, which means that each consecutive configuration can be considered as a new equilibrium. Therefore, using the Kuperus-Raadu model of prominence equilibrium one can write (Kuperus and Raadu \cite{Kuperus1974}, Priest \cite{Priest1982})
\begin{equation}
{B^2\over {4 \pi}}=\rho g h
\end{equation}
where $B$ is the magnetic field strength, $\rho$ is the plasma density, $g$ is the surface gravitational acceleration and $h$ is the height from the surface. During the slow rise phase, the magnetic field does not change significantly (as also showed by magnetic field extrapolation). Then the slight increase of the prominence height must be balanced by slight decrease of the prominence density. Figure 15 shows the observed dynamics of prominence height and estimated decrease of density due to coronal rain during the whole interval. It is clearly seen that the height and density are in anti-phase; one increases and the other decreases. We also see that the slow rise of prominence is linear indicating that the cause of the rising is not instability (see the exponential character in fast rise phase). This rough estimation supports the suggestion that the slow rise of prominence is caused by decreasing of prominence mass due to coronal rain blobs.

\section{Discussion}

We analysed three solar prominences observed during 2011-2012 using simultaneous observations of SDO and Stereo spacecrafts. The prominences were observed from different angles by different missions, which gave us possibility to follow the dynamics of prominences in details.

In all three cases, the massive coronal rain blobs started to flow from the prominence main bodies and after few tenth of hours prominences destabilised and erupted as CMEs. It is shown that after few hours of coronal rain prominences started to rise up probably due to the mass loss. The rise of prominence consists in two phase: the initial slow rise phase and the consecutive fast rise phase.

\begin{figure}
\begin{center}
\includegraphics[width=6cm]{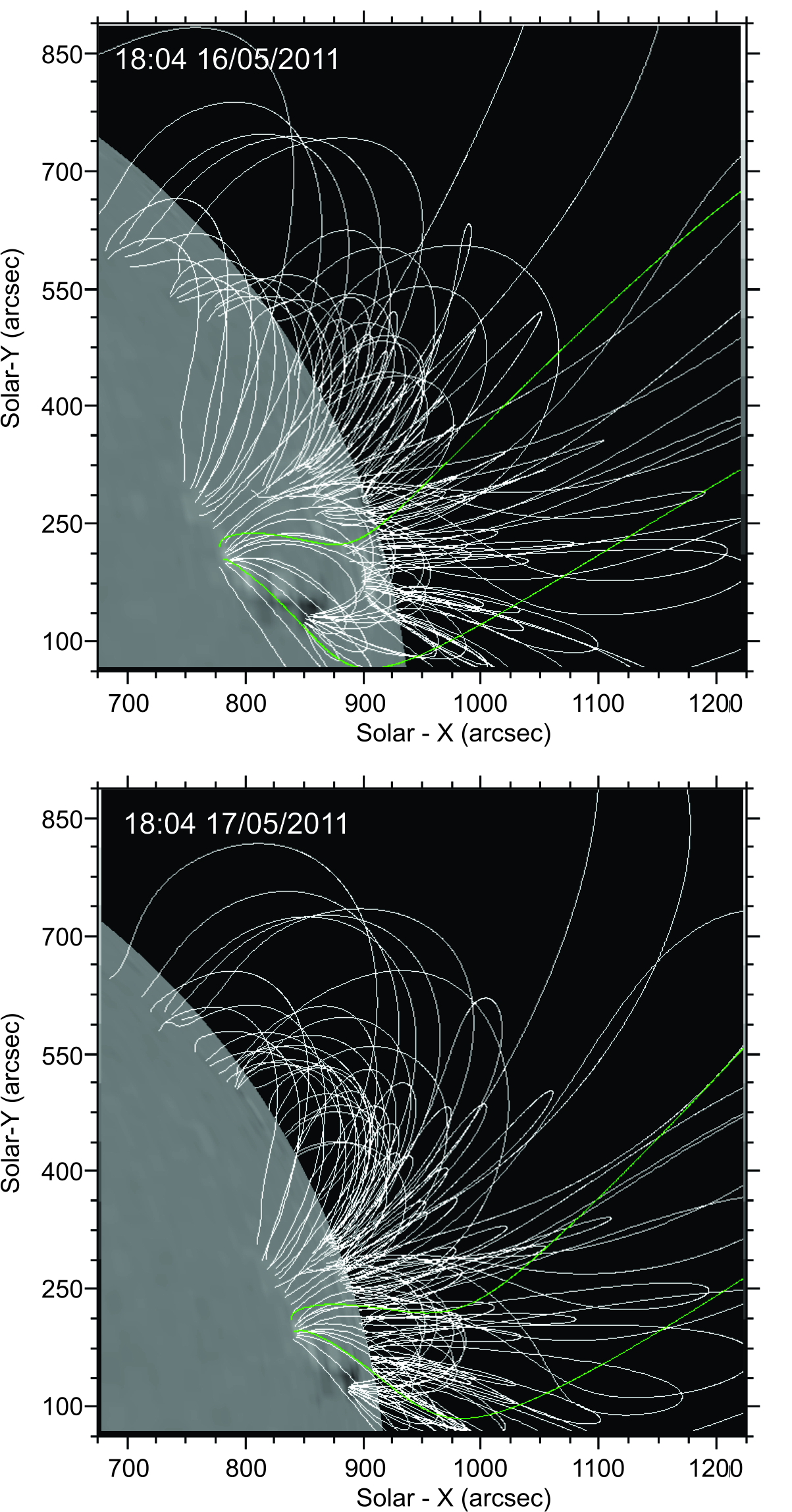}
\end{center}
\caption{Magnetic field extrapolation (potential-field source-surface PFSS) in prominence area before (at 18:04 UT, May 16, upper panel) and during (at 18:04 UT, May 17, lower panel) the slow rise process. White curves correspond to the close magnetic field lines, green (purple) lines show the open magnetic field lines with positive (negative) polarity.}
\end{figure}

We tracked falling coronal rain blobs along inclined trajectories (totally 32 trajectories were detected in the three cases). From tracked trajectories, we extracted the space-time diagram to define acceleration and initial velocity of the falling plasma blobs. We fitted well-defined curved trajectories from space-time diagrams with polynomial fits and estimated initial velocity with a main around 65 km s$^{-1}$, which correspond to the values reported in previous works (Schrijver \cite{Schrijver2001}, De Groof et al. \cite{DeGroof2005}, Antolin \cite{Antolin2010}, Antolin \& Verwichte \cite{Antolin2011}, Antolin \& Rouppe van der Voort \cite{Antolin2012}).

Accelerations of coronal rain blobs were estimated as 20-136 m s$^{-2}$ with a mean around 74 m s$^{-2}$ from all tracked coronal rain blobs. The acceleration is smaller than the free fall as found  in earlier works (Schrijver \cite{Schrijver2001}, De Groof et al. \cite{DeGroof2004}, De Groof et al. \cite{DeGroof2005}, Antolin \cite{Antolin2010}, Antolin \& Verwichte \cite{Antolin2011}, Antolin \& Rouppe van der Voort \cite{Antolin2012}). The falling distance were more than 29 $Mm$ for most of the blobs and falling time was $>$ 30 min as reported in previous works ( De Groof et al. \cite{DeGroof2004}, De Groof et al. \cite{DeGroof2005}, Antolin \cite{Antolin2010}, Antolin \& Verwichte \cite{Antolin2011}, Antolin \& Rouppe van der Voort \cite{Antolin2012}).

We estimated the prominence mass loss by coronal rain blobs before destabilisation as around 17-19 \%. However, in all three cases we observed the coronal rain blobs only along one leg of the prominences. Another legs were not seen by observations. If one assumes the symmetric mass flow along the both legs, then the mass loss will increase to 34-38 \%. Therefore, one may conclude that the instability starts when prominences loose around 40 \% of their masses. If initial density in the considered prominences was as $\rho$=1.67 $\times$10$^{-14}$ g cm$^{-3}$, then after the mass loss due to coronal rain the density may become as (considering unchanged volumes of prominence bodies) 1.11 $\times$10$^{-14}$ g cm$^{-3}$ (for the event of May 16-18, 2011), 1.02 $\times$10$^{-14}$ g cm$^{-3}$ (for the event of December 22-24, 2011) and  1.12 $\times$10$^{-14}$ g cm$^{-3}$ (for the event of August 07-08, 2012).

\begin{figure}
\begin{center}
\includegraphics[width=\columnwidth]{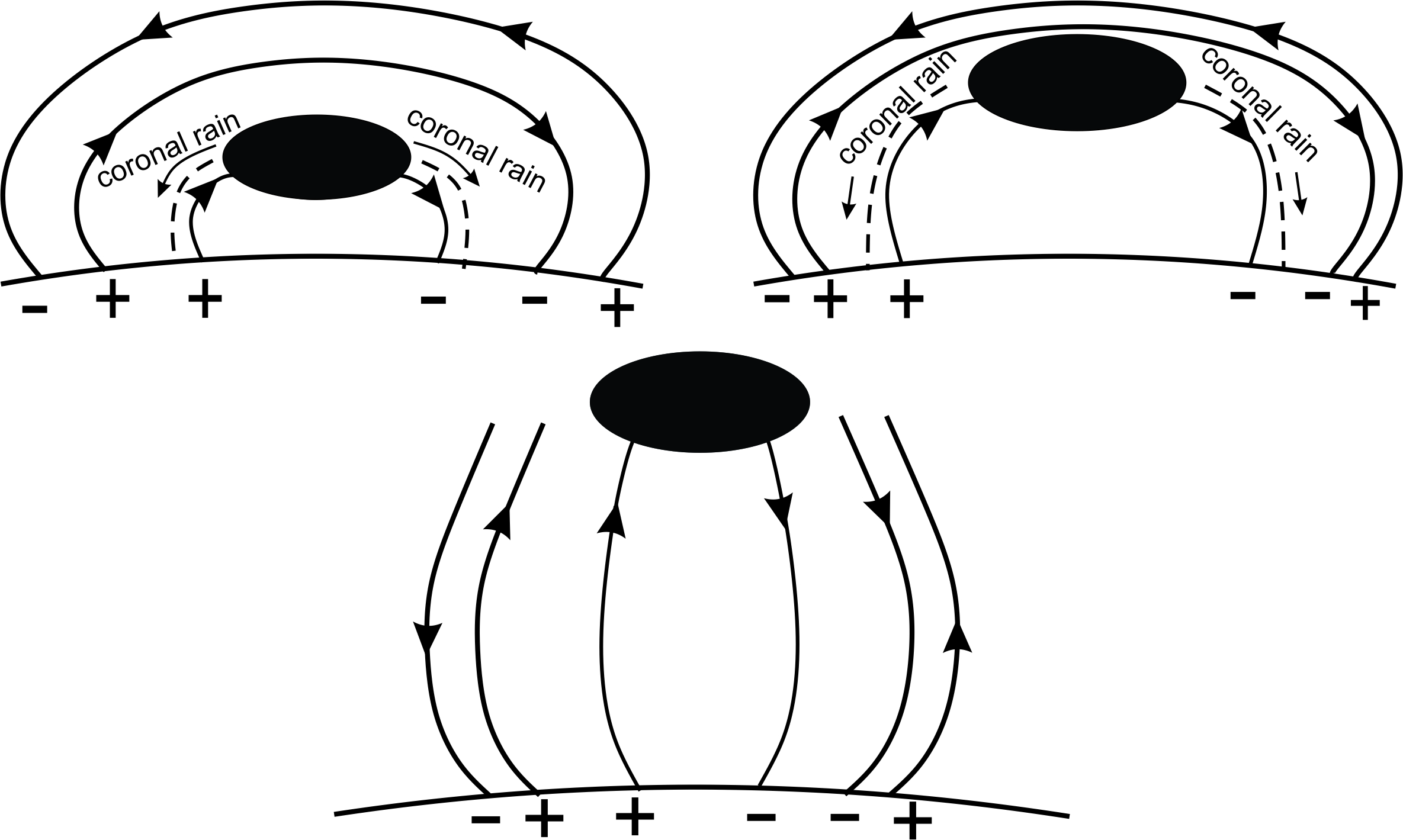}
\end{center}
\caption{Schematic dynamics of prominence evolution according to our model. Black ellipse represents the prominence/filament and dashed curved lines show the coronal rain blob trajectories. Initial configuration is shown by  the upper left panel.  Upper right panel displays the slow rise phase of the prominence triggered by the coronal rain. Final instability during fast rise phase and consecutive eruption of prominence as CME is shown on the lower panel.}
\end{figure}

To understand the triggering mechanism for the initial slow rise of prominence is of vital importance. The mechanism could be related with unstable magnetic field configuration e. g. the kink instability of twisted flux tubes. The growth time for the kink instability in prominence conditions can be estimated as $<$ 1 hour (see previous section). But, the slow rise phase lasts much longer in all three cases; it is around 25 hours for the event of May 16-18 (2011), for example. Therefore, the kink instability can be ruled out as a triggering mechanism for the slow rise phase. On the other hand, reduction of density and mass in prominences due to the coronal rain obviously causes the reduction of gravity force, which eventually led to the excess of Lorentz force and hence upward motion of prominence bodies. We studied the detailed upward motion of the prominence using space-time diagram for the event of May 16-18 (2011), which showed that the slow rise phase could be triggered by coronal rain. The slow rise of the prominence at some height may eventually lead to the torus instability (T\"or\"ok and Kliem 2007, Filippov \cite{Filippov2013}, Zuccarello et al. \cite{Zuccarello2014}) or reconnection with overlying coronal magnetic field according to the breakout model (Antiochos et al. 1999). The instabilities correspond to the fast rise phase, when prominences finally erupted as CMEs (see Figure 14).

\begin{figure}[t]
\vspace*{1mm}
\begin{center}
\includegraphics[width=\columnwidth]{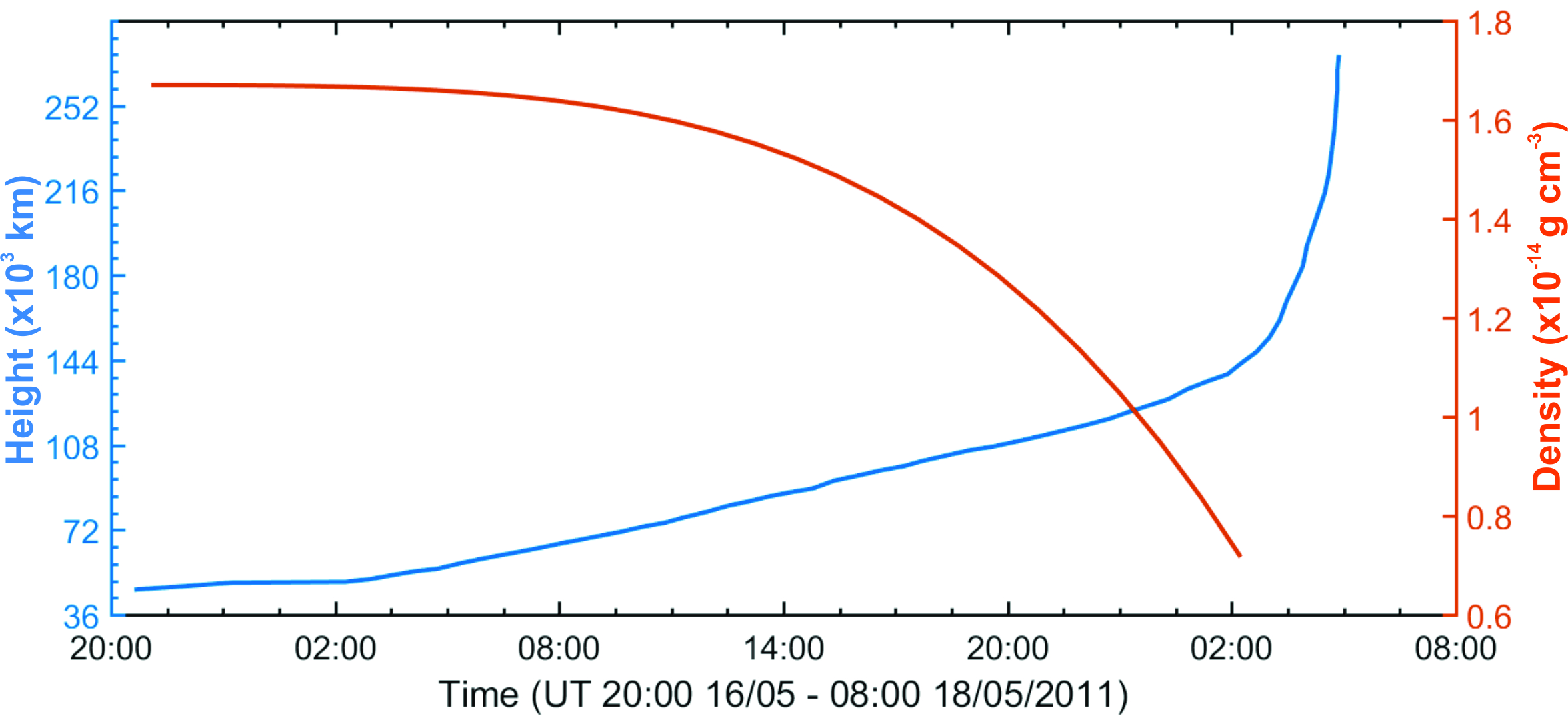}
\end{center}
\caption{Upward rise of the prominence along the cut on Figure 4 (blue line) and estimated decrease of density due to coronal rain (red line) during the whole evolution of prominence in the event of May 16-18, 2011.}
\end{figure}

The time intervals between the start of coronal rain and final destabilisation of the prominence were estimated as 28 hours for the event of May 16-18 (2011), 26 hours for the event of December 22-24 (2011), and 18 hours for the event of August 07-08 (2012). Therefore, if the coronal rain leads to the initial slow rise and the consequent final destabilisation of many prominences, then one can predict CMEs 20-30 hours before their actual eruptions. This will lead to considerable improvement of space weather predictions. This requires detailed statistical study of interconnection between coronal rain and CME initiations.

\section{Conclusion}

We studied the dynamics of three different prominences/filaments (May 16-18, 2011, December 22-24, 2011, and August 07-08, 2012) using simultaneous observational data from SDO/AIA and Stereo/SECCHI. SDO and Stereo twin spacecrafts were observing the Sun from different angles, therefore we could trace the evolution of the prominences in detailed. In all three cases, the massive coronal rain started to fall down towards the photosphere, which followed by the destabilisation of prominences after 20-30 hours and later by CMEs. The upward rise of prominence consisted of two phases: initial slow rise phase and consecutive fast rise phase. We suggest that the slow rise phase was triggered by the violation of equilibrium between gravity and Lorentz forces as the result of the mass loss due to coronal rain. On the other hand, the fast rise phase was probably connected with magnetic instability/reconnection, which led to the final destabilisation of the prominences. Estimations of mass flux along one legs of the prominences (another legs were not seen by observations) led to 20 \% of the mass loss before the final destabilisation. Assuming the symmetric mass flux along the both legs, one can conclude that the prominences became unstable after loss of 40 \% of their masses. If future analysis show the similar behaviour for many prominences then the coronal rain may be used to predict the prominence instability and hence CMEs. This will help to improve the space weather predictions.

\begin{acknowledgements}
Work was supported by the Shota Rustaveli National Science Foundation (SRNSF) grant DI-2016-52, The work  of TVZ and AH was funded by the Austrian Science Fund (FWF, projects P30695-N27 and I 3955-N27).
PG acknowledges the support of the project VEGA 2/0048/20,
We thank the anonymous referee for useful comments, which led to improving the paper significantly.
\end{acknowledgements}

\appendix

\end{document}